\documentclass[aps,pra,twocolumn,superscriptaddress]{revtex4-2}
\usepackage[utf8]{inputenc}

\date{\today}
\usepackage{amsmath}
\usepackage{amssymb}
\usepackage{mathtools}
\usepackage{caption}
\usepackage{subcaption}
\usepackage[colorlinks]{hyperref}
\DeclarePairedDelimiter{\abs}{\lvert}{\rvert}%
\DeclarePairedDelimiterX\innerp[2]{\langle}{\rangle}{#1\delimsize\vert\mathopen{}#2}%
\DeclarePairedDelimiterX\braket[2]{\langle}{\rangle}{#1\delimsize\vert\mathopen{}#2}%
\DeclarePairedDelimiterX\braketOP[3]{\langle}{\rangle}{#1\,\delimsize\vert\,\mathopen{}#2\,\delimsize\vert\,\mathopen{}#3}%
\DeclarePairedDelimiterX\ketbra[2]{\lvert}{\rvert}{#1\delimsize\rangle\!\delimsize\langle#2}%
\DeclarePairedDelimiterX\outerp[2]{\lvert}{\rvert}{#1\delimsize\rangle\!\delimsize\langle#2}%
\DeclarePairedDelimiterX\projector[1]{\lvert}{\rvert}{#1\delimsize\rangle\!\delimsize\langle#1}%
\DeclareMathOperator{\tr}{tr}%

\begin{document}
\title{Recycling Reflections for Perfect Photon Capture}

\author{Yat Wong}
\affiliation{Pritzker School of Molecular Engineering, University of Chicago, Chicago, Illinois 60637, USA}
\author{Liang Jiang}
\affiliation{Pritzker School of Molecular Engineering, University of Chicago, Chicago, Illinois 60637, USA}
\begin{abstract}
Efficient photon capture in optical cavities is essential for quantum networks and computing, yet single-pass methods suffer from uncaptured reflections due to finite capture windows and coupling strengths, precluding perfect transfer of arbitrary photonic states. We introduce a two-pass ‘pitch-and-catch’ method that recycles initial reflections to achieve perfect capture of arbitrary pulses in the noiseless regime and enhances fidelity under intrinsic losses. The method extends to photon emission, enabling arbitrary pulse shaping. This advance offers significant improvements for quantum repeaters, memories, and transduction, enhancing the toolkit for quantum information processing.
\end{abstract}
\maketitle
\section{Introduction}
Capturing and emitting flying photonic states are cornerstones of quantum optics, enabling state storage, transfer, transduction, and manipulation in quantum networks and computing platforms~\cite{Kimble_2008,Lauk_2020}. Early theoretical work was established by Cirac et al., proposing quantum state transfer between distant nodes via photon capture in cavities, laying the groundwork for quantum networks~\cite{Cirac_1997}. Experimental advances have demonstrated quantum state transfers between matter and light~\cite{Matsukevich_2004,Boozer_2007,Stute_2013,Axline_2018}. When given infinite capture time, an input shaped to an exponential growth pulse can be captured perfectly~\cite{Stobinska_2009}. In practice, practical constraints results in uncaptured reflections that degrade fidelity due to finite capture windows and coupling strengths, even with control of both sending and receiving nodes~\cite{Cirac_1997}, and hence, under such constraints, a perfect transfer of arbitrary single mode state is impossible with a single pass. Similarly, exponential decay pulses cannot be captured with finite coupling strength even in the asymptotic limit due to initial reflections.

To overcome these challenges, we propose a two-pass ‘pitch-and-catch’ photon capture method that recycles initial reflections to \emph{perfectly} transfer arbitrary pulse in the noiseless regime for exponential decay pulses in the asymptotic limit and finite time frame pulses, and increase the fidelity in the presence of noise. The method can also be used for photon emission, utilizing the first pass's emission to extract the remaining excitation in the second pass within a finite time frame. Section 2 formulates the problem using the quantum Langevin equation. Section 3 derives the theoretical limit of single-pass capture, while Section 4 introduces the two-pass method for arbitrary pulses. Sections 5 and 6 extend the analysis to exponential decay pulses in noiseless and noisy regimes, respectively.
\section{Methodology}

\begin{figure}[]
\begin{subfigure}[b]{.4\textwidth}
\includegraphics[width=\textwidth]{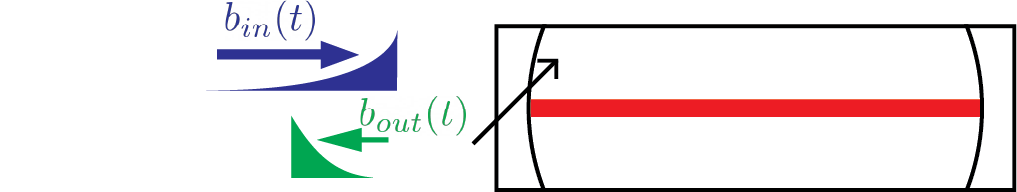}
\caption{In existing schemes, the input pulse only interacts with the cavity once, leaving a reflection uncaptured.}
\end{subfigure}
\begin{subfigure}[b]{.4\textwidth}
\includegraphics[width=\textwidth]{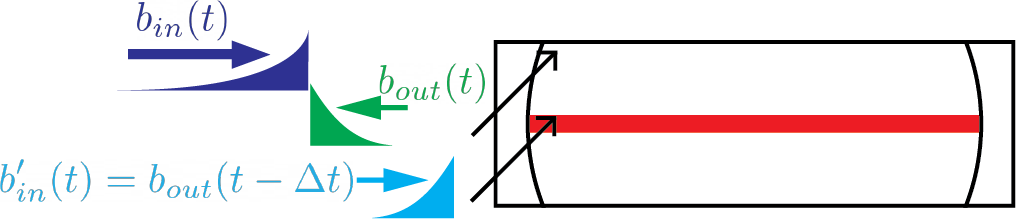}
\caption{In our proposed scheme, the reflection is redirected back to the cavity after a delay \(\Delta t\) for a second pass of capture.}
\end{subfigure}
\caption{Multiple-pass photon capture} \label{fig:passes}
\end{figure}
We model photon capture using a quantum Langevin equation for a cavity mode annihilation operator \(\hat{a}(t)\) coupled to a traveling mode input \(\hat{b}_{in}(t)\) via a tunable coupling \(\kappa(t)\), and we analyze the problem in the Heisenberg picture. We assume the cavity starts in a vacuum state, the input state is single-mode, and the system either operates in a linear regime or has at most one excitation, allowing dynamics to be captured by the single-photon wavefunction or equivalently classical amplitudes. Note that such cavity is a visualization of the receiving bosonic mode, and is assumed to be unimodal and in the high-finesse limit. This process can also be applied to non-bosonic devices, such as atomic ensembles in the linear regime, and qubits or atom-cavity systems~\cite{Cirac_1997} when the input state has no more than one excitation. The input state is represented by a lowering operator \(\hat{c}\), a superposition of \(\hat{b}_{in}(t)\), and a corresponding density operator 
\begin{equation}
\rho_{in}:=\sum_{nm}\frac{\rho_{nm}}{\sqrt{n!m!}}\hat{c}^{\dagger n}\projector{0}\hat{c}^m,
\end{equation}
where \(\rho_{nm}\) are the density matrix elements defining the input state. The lowering operator of the cavity obeys the following equation:
\begin{equation}
\frac{d}{dt}\hat{a}(t)=-\frac{\kappa(t)}{2}\hat{a}(t)+\sqrt{\kappa(t)}\hat{b}_{in}(t).
\end{equation}
In this regime, the process corresponds to a beamsplitter operation between the cavity and input mode, and \(\hat{a}(t)\) is a superposition of \(\hat{a}(0)\), \(\hat{c}\), and a superposition of \(\hat{b}(t')\) orthogonal to \(\hat{c}\). Since \(\hat{a}(0)\) starts in the vacuum state and the input is unimodal, the commutator \([\hat{a}(t), \hat{c}^\dagger]=a(t)\) is a scalar, which can be interpreted as the complex amplitude of \(\hat{a}(t)\) projected in \(\hat{c}\). This complex amplitude completely quantifies the efficiency of this capture process, since we have \(\left[\hat{c}-a^*(t)\hat{a}(t),\hat{a}^\dagger(t)\right]=a^*(t)-a^*(t)=0\), so if we let $\hat{c}-a^*(t)\hat{a}(t)=\sqrt{1-\abs{a}^2}\hat{a}'(t)$, where \(\hat{a}'(t)\) is orthogonal to \(\hat{c}\) and is a superposition of \(\hat{b}_{in}(t'>t)\) and \(\hat{b}_{out}(t'<t)\), then
\begin{equation}
\rho_a(t)=\tr_{\hat{a}'(t)}\left(\sum_{nm}\frac{\rho_{nm}}{\sqrt{n!m!}}\hat{c}^{\dagger n}\projector{0}\hat{c}^m\right),
\end{equation}
as \(\hat{a}(0)\) started with vacuum and the input is unimodal. Hence, we only need to track the commutator between the relevant lowering operators and the raising operator of the input pulse, e.g. $b_{in}(t)=\left[\hat{b}_{in}(t),\hat{c}^\dagger\right]$. \(b_{in}(t)\) is the input pulse shape, i.e. \(\hat{c}^\dagger=\int b_{in}(t)\hat{b}_{in}^\dagger(t)dt\), but here we interpret it as the complex amplitude of \(\hat{b}_{in}(t)\) projected in \(\hat{c}\). Note that the \(t\) dependence here does not arise from time-dependent dynamics of a mode like \(a(t)\) from \(\hat{a}(t)\), but a time-dependent label of an infinite family of orthogonal modes \(\hat{b}_{in}(t)\). The problem is then reduced to optimizing \(\kappa(t)\) to capture an input photon represented by the normalized pulse $b_{in}(t)$, and this work focuses on extending the analysis to two-pass configurations with delayed reflection (see FIG. \ref{fig:passes}). Such cavity obeys the following equation:
\begin{equation}
\begin{split}
\frac{d}{dt}a(t)=&-\frac{\kappa_1(t)}{2}a(t)+\sqrt{\kappa_1(t)}b_{in}(t)\\
&-\frac{\kappa_2(t)}{2}a(t)+\sqrt{\kappa_2(t)}b'_{in}(t),
\end{split}
\end{equation}
where
\begin{equation}
b_{in}'(t)=b_{out}(t-\Delta t)=b_{in}(t-\Delta t)-\sqrt{\kappa_1(t-\Delta t)}a(t-\Delta t),
\end{equation}
is the redirected reflection and \(\Delta t\) is the delay for the reflection to reach the cavity again, which we assume to be controllable. Note that \(\hat{b}'_{in}\) is in a mode orthogonal to \(\hat{b}_{in}\), and \(\kappa_2\) is controllable independently of \(\kappa_1\).
\section{Single Pass Limit}

\begin{figure}[]
\includegraphics[width=0.45\textwidth]{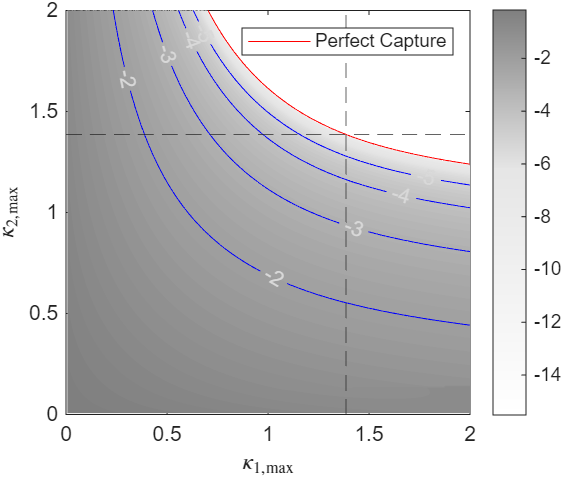}
\caption{Logarithm of minimum loss achievable with two-pass capture for square pulse \(\log_{10}(1-a^2_{\max})\), where \(a_{\max}=\max_{\kappa_1,\kappa_2}a(t=2b_{\max}^{-2})\). White region corresponds to perfect capture. Dashed line indicates \(\kappa=2\ln{2}\), the minimum \(\kappa_{\max}\) required to achieve perfect capture when \(\kappa_{1,\max}=\kappa_{2,\max}=\kappa_{\max}\). \(\kappa_{1,\max}\) and \(\kappa_{2,\max}\) are in the units of \(b_{\max}^2\).}
\label{fig:square_loss}
\end{figure}

\begin{figure}[]
\begin{subfigure}[b]{0.35\textwidth}
\includegraphics[width=\textwidth]{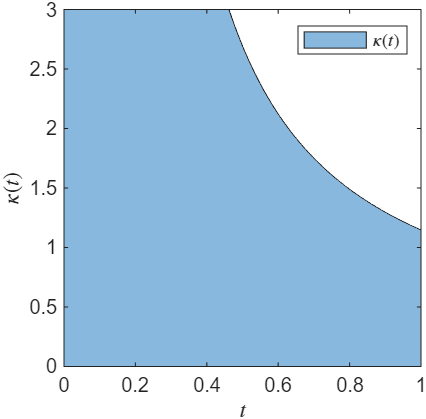}
\caption{Single-pass \(\kappa(t)\)}
\end{subfigure}
\begin{subfigure}[b]{0.35\textwidth}
\includegraphics[width=\textwidth]{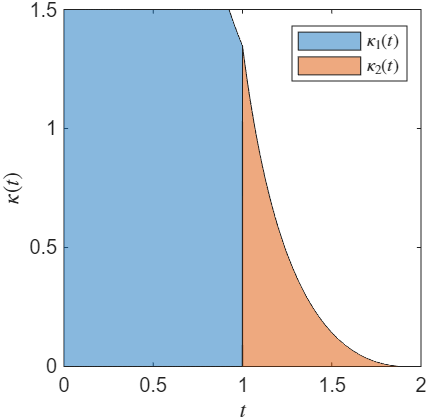}
\caption{Two-pass \(\kappa(t)\)}
\end{subfigure}
\begin{subfigure}[b]{0.35\textwidth}
\includegraphics[width=\textwidth]{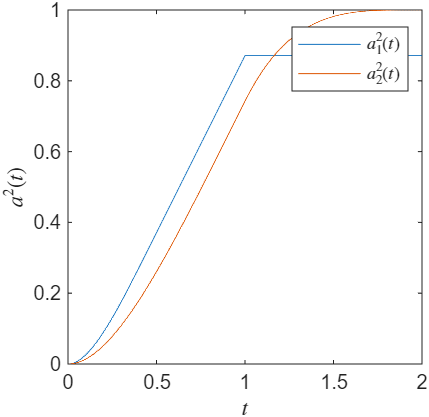}
\caption{\(a^2(t)\) for both schemes}
\end{subfigure}
\caption{Example pulse of single-pass with \(\kappa_{\max}=3b_{\max}^2\) and  two-pass capture with \(\kappa_{1,\max}=\kappa_{2,\max}=1.5b_{\max}^2\), and the corresponding efficiency \(a^2(t)\).\(\kappa(t),\kappa_{1}(t)\) and \(\kappa_{2}(t)\) are in the units of \(b_{\max}^2\), while \(t\) is in the units of \(b_{\max}^{-2}\). }
\label{fig:square_pulse}
\end{figure}
Given a cavity mode $a$ and input pulse $b_{in}$, the dynamics of the cavity mode is as follows:
\begin{equation}
\frac{da}{dt}=-\frac{\kappa}{2}a+\sqrt{\kappa}b_{in},
\end{equation}
where $t$ is time and $\kappa$ is the coupling strength. We want to capture a pulse, which we assume that can be shaped arbitrarily, with a maximal coupling strength $\kappa_{\max}$ within a capture window $t_{\max}$. We reparametrize the equation with \(\beta=\kappa^{-1/2}b_{in}\) and \(\tau=\int_0^tdt'\kappa(t')\) to simplify the dynamics. Hence $\tau_{\max}=\int_0^{t_{\max}}dt'\kappa(t')\leq\kappa_{\max}t_{\max}$. The resultant equation is
\begin{equation}
\frac{da}{d\tau}=-\frac{1}{2}a+\beta,
\end{equation}
and solving it results in
\begin{equation}
a(\tau_{\max})=\int_0^{\tau_{\max}}d\tau e^{(\tau-\tau_{\max})/2}\beta(\tau).
\end{equation}
Hence, the change of \(a(\tau_{\max})\) under variation of \(\beta(\tau)\) is given by the following equation:
\begin{equation}\label{eqa}
\delta a(\tau_{\max})=\int_0^{\tau_{\max}}d\tau e^{(\tau-\tau_{\max})/2}\delta\beta(\tau),
\end{equation}
while the normalization varies as follows:
\begin{equation}\label{eqb}
\delta\int_0^{\tau_{\max}}d\tau\beta(\tau)^2=2\int_0^{\tau_{\max}}d\tau\beta(\tau)\delta\beta(\tau).
\end{equation}
Therefore, to maximize $a(\tau_{\max})$ after the capture under the constraint of \(\int_0^{\tau_{\max}}\beta(\tau)^2d\tau=1\), we use Lagrange multipliers, yielding \(\beta(\tau)\propto e^{\tau/2}\). Applying normalization gives
\begin{equation}
\beta(\tau)=\frac{e^{\tau/2}}{\sqrt{e^{\tau_{\max}}-1}},
\end{equation}
and the maximal amplitude of $a$ would be
\begin{equation}
a(\tau_{\max})=\int_0^{\tau_{\max}}d\tau\frac{e^{\tau-(\tau_{\max}/2)}}{\sqrt{e^{\tau_{\max}}-1}}=\sqrt{1-e^{-\tau_{\max}}}.
\end{equation}
The maximum amplitude is thus $a(t_{\max})=\sqrt{1-e^{-\kappa_{\max}t_{\max}}}$, achieved with an exponential growth pulse \(b_{in}(t)=\frac{\sqrt{\kappa_{\max}}e^{\kappa_{\max}t/2}}{\sqrt{e^{\kappa_{\max}t_{\max}}-1}}\) and constant \(\kappa(t)=\kappa_{\max}\). However, with finite \(t_{\max}\) and \(\kappa_{\max}\), residual reflection \(e^{-\kappa_{\max}t_{\max}}\) is inevitable for any single-pass process.
\section{Two-pass capture of arbitrary positive packet with bounded magnitude}
As shown in Section 3, single-pass capture leaves a residual reflection \(e^{-\kappa_{\max}t_{\max}}\), which we aim to eliminate using a two-pass approach, i.e. allowing the reflection to interact with the cavity again. Such a system would follow a Langevin equation in this form:
\begin{equation}
\frac{d}{dt}a(t)=-\frac{\kappa_1(t)+\kappa_2(t)}{2}a(t)+\sqrt{\kappa_1(t)}b_{in}(t)+\sqrt{\kappa_2(t)}b_{in}'(t).
\end{equation}
Here, \(\kappa_1(t)\) and \(\kappa_2(t)\) are the tunable couplings for the first and second ports, and \(b_{in}'(t)=b_{out}(t-\Delta t)\) is the redirected reflection.
In this section, we assume the input pulse is positive and bounded, i.e. $0<b_{in}\left(t\right)<b_{\max}$ for any time $t$ during the packet. Note that if one can tune the phase of coupling, complex pulses can be captured with similar processes. Here we choose the time unit that satisfies $b_{\max}=1$, i.e. \(t\) will be in the units of \(b_{\max}^{-2}\) and \(\kappa\) will be in the units of \(b_{\max}^2\). Consider the instantaneous energy capture efficiency at time $t$,
\begin{equation}
\begin{split}
\frac{\frac{d}{dt}a^2(t)}{b_{in}^2\left(t\right)}&=-\kappa_1(t)\frac{a^2(t)}{b_{in}^2\left(t\right)}+2\sqrt{\kappa_1(t)}\frac{a(t)}{b_{in}\left(t\right)}\\&=1-\left(1-\sqrt{\kappa_1(t)}\frac{a(t)}{b_{in}\left(t\right)}\right)^2
\end{split}
\end{equation}
\begin{equation}\label{kappa1}
\kappa_1(t)=\min\left(
\kappa_{1,\max}, \left(\frac{b_{in}\left(t\right)}{a(t)}\right)^2\right)
\end{equation}
We can see that this choice corresponds to eliminate reflection if possible, and maximize $\kappa_1$ otherwise. The square packet, i.e. $\hat{c}=\int_0^1\hat{b}_{in}\left(t\right)dt$, maximizes initial reflection, making it the most challenging case for complete capture. For \(\kappa_{1,\max}\leq2\ln{2}\), \(\kappa_1\) is held at \(\kappa_{1,\max}\) throughout the packet, yielding
\begin{equation}
a^2\left(t=1\right)\bigg\vert_{\kappa_{1,\max}\leq2\ln{2}}=\frac{4}{\kappa_{1,\max}}\left(1-e^{-\kappa_{1,\max}/2}\right)^2,
\end{equation}
while for \(\kappa_{1,\max}>2\ln{2}\), \(\kappa_1\) is reduced to \(\frac{\kappa_{1,\max}}{1+\kappa_{1,\max}(t-t_1)}\) after \(t>t_1=\frac{2\ln2}{\kappa_{1,\max}}\), at which the capture process shifts into a reflection-less regime, resulting in
\begin{equation}
a^2\left(t=1\right)\bigg\vert_{\kappa_{1,\max}>2\ln{2}}=1-\frac{2\ln{2}-1}{\kappa_{1,\max}}.
\end{equation}\(\)
In the second pass, similar to the first pass, \(\kappa_2\) would be set to
\begin{equation}\label{kappa2}
\kappa_2(t)=\min\left(
\kappa_{2,\max}, \left(\frac{b'_{in}\left(t\right)}{a(t)}\right)^2\right)
\end{equation}
The minimal $\kappa_{2,\max}$ required to completely capture the reflected wavepacket from the first pass would then be
\begin{equation}
\kappa_{2,\max}\geq\begin{cases}
\frac{\kappa_{1,\max}}{4\left(1-e^{-\kappa_{1,\max}/2}\right)^2}, \text{ if }\kappa_{1,\max}\leq2\ln{2}\\
\left(1-\frac{2\ln{2}-1}{\kappa_{1,\max}}\right)^{-1}, \text{ if }\kappa_{1,\max}>2\ln{2}
\end{cases}
\end{equation}
The achievable loss is illustrated by FIG. \ref{fig:square_loss}. If $\kappa_{1,\max}=\kappa_{2,\max}=\kappa_{\max}$, the minimal $\kappa_{\max}$ required is $2\ln{2}$. Hence, for any bounded packet, \(\kappa_{\max}\geq2\ln{2}\) would guarantee perfect two-pass capture, while single-pass efficiency could be limited to \(1-O(\kappa_{\max}^{-1})\), demonstrated by FIG. \ref{fig:square_pulse}. Note that in this section, the two passes are performed separately for better visualization, but in principle they can be performed concurrently; for a square pulse of length \(T\), the two-pass capture can be completed within the time frame of the pulse if \(\kappa_{\max}\geq\frac{4\ln{2}}{T}\).
\section{Exponential decay packet}
\begin{figure}[]
\includegraphics[width=0.45\textwidth]{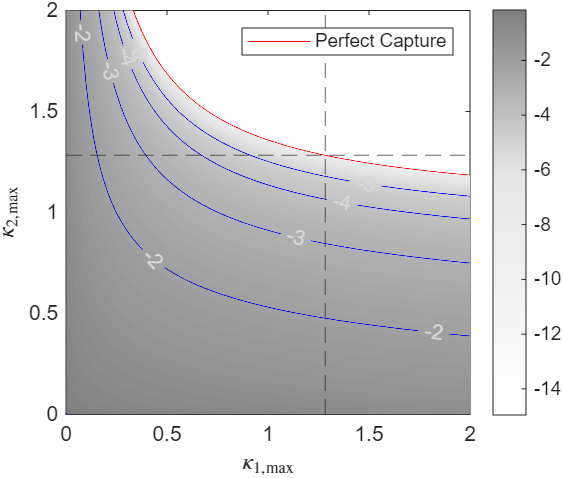}
\caption{Logarithm of minimum loss achievable with two-pass capture for exponential decay pulse \(\log_{10}(1-a^2_{\sup})\), where \(a_{\sup}=\sup_{\kappa_1,\kappa_2}\lim_{t\rightarrow\infty}a(t)\). White region corresponds to perfect capture. Dashed line indicates \(\kappa=k\), the minimum \(\kappa_{\max}\) required to achieve perfect capture when \(\kappa_{1,\max}=\kappa_{2,\max}=\kappa_{\max}\). \(\kappa_{1,\max}\) and \(\kappa_{2,\max}\) are in the units of \(\gamma\). }
\label{fig:noiseless_loss}
\end{figure}

\begin{figure}[]
\begin{subfigure}[b]{0.35\textwidth}
\includegraphics[width=\textwidth]{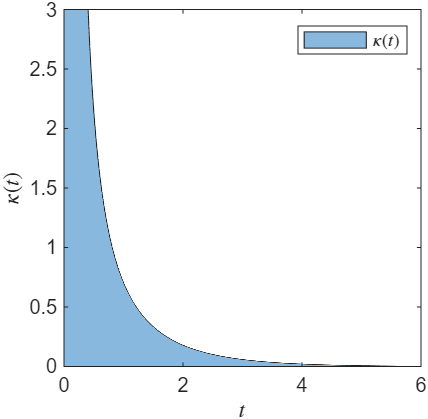}
\caption{Single-pass \(\kappa(t)\)}
\end{subfigure}
\begin{subfigure}[b]{0.35\textwidth}
\includegraphics[width=\textwidth]{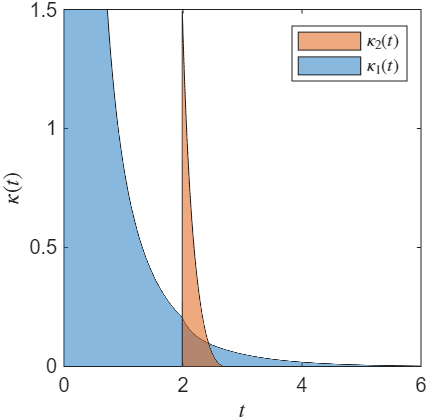}
\caption{Two-pass \(\kappa(t)\)}
\end{subfigure}
\begin{subfigure}[b]{0.35\textwidth}
\includegraphics[width=\textwidth]{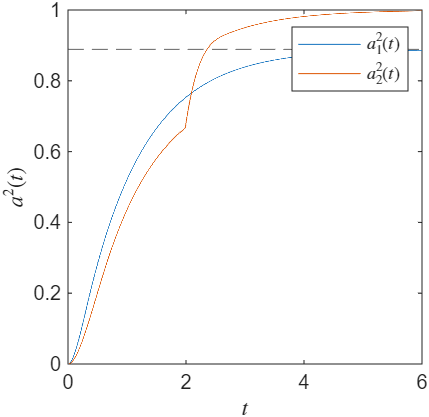}
\caption{\(a^2(t)\) for both schemes}
\end{subfigure}
\caption{Example pulse of single-pass with \(\kappa_{\max}=3\gamma\) and two-pass capture with \(\kappa_{1,\max}=\kappa_{2,\max}=1.5\gamma\), and the corresponding efficiency \(a^2(t)\). Dashed line indicates the maximum efficiency of the single-pass scheme in the asymptotic limit. \(\kappa(t),\kappa_{1}(t)\) and \(\kappa_{2}(t)\) are in the units of \(\gamma\), while \(t\) is in the units of \(\gamma^{-1}\). }
\label{fig:noiseless_pulse}
\end{figure}
Exponential decay pulses, common in quantum optics due to spontaneous emission, provide a practical case study for our method. Similar to the previous section, we choose time unit that satisfies $\gamma=1$, i.e. \(t\) will be in the units of \(\gamma^{-1}\) and \(\kappa\) will be in the units of \(\gamma\), and assume zero intrinsic loss $ \kappa_{i}=0$. For the exponential decay packet
\begin{equation}
b_{in}\left(t\right)=e^{-t/2},
\end{equation}
equation \ref{kappa1} gives a similar choice of \(\kappa_1(t)\) similar to previous section: initially, \(\kappa_1(t)=\kappa_{1,\max}\) and reduced to \(\frac{\kappa_{1,\max}e^{-t+t_1}}{1+\kappa_{1,\max}\left(1-e^{-t+t_1}\right)}\) for \(t>t_1=\frac{2}{\kappa_{1,\max}-1}\ln{\frac{2\kappa_{1,\max}}{\kappa_{1,\max}+1}}\), resulting in 
\begin{equation}
a_1\left(t\geq t_1\right)=e^{-t_1/2}\sqrt{\frac{1}{\kappa_{1,\max}}+1-e^{-t+t_1}}.
\end{equation}
With a single pass, the maximal capture efficiency is thus 
\begin{equation}
a^2_1(t\rightarrow\infty)=\left(\left(1+\kappa_{1,\max}^{-1}\right)^{\kappa_{1,\max}+1}/4\right)^{\left(\kappa_{1,\max}-1\right)^{-1}},
\end{equation}
which is approximately \(1-\frac{2\ln{2}-1}{\kappa_{1,\max}}\) for large \(\kappa_{\max}\). To make $b'_{out}\left(t\right)=0$, it is clear that the maximal $\kappa_2$ is only required at $t=\Delta t$, thus
\begin{equation}
1-\sqrt{\kappa_{2,\max}}a\left(\Delta t\right)=b'_{out}\left(\Delta t\right)=0.
\end{equation}
Therefore, to achieve perfect capture, we must have
\begin{equation}
\kappa_{2,\max}\left(\left(1+\kappa_{1,\max}^{-1}\right)^{\kappa_{1,\max}+1}/4\right)^{\left(\kappa_{1,\max}-1\right)^{-1}}>1.
\end{equation}
Assuming \(\Delta t\geq t_1\), we have
\begin{equation}
\sqrt{1+\kappa_{1,\max}\left(1-e^{-\Delta t+t_1}\right)}=\sqrt{\frac{\kappa_{1,\max}}{\kappa_{2,\max}}}e^{t_1/2}
\end{equation}
\begin{equation}
\Delta t=t_1+\ln{\frac{1}{1+\kappa_{1,\max}^{-1}-\kappa_{2,\max}^{-1}e^{t_1}}}.
\end{equation}
In other words, if \(\kappa_{2,\max}>\kappa_{1,\max}e^{t_1}\), one can start capturing the reflection before the first pass reaches reflectionless regime. If \(\kappa_{1,\max}=\kappa_{2,\max}=\kappa_{\max}\), the infimal \(\kappa_{\max}\) required to achieve perfect capture is $\kappa_{\max}>k\approx 1.2834$, where $k$ satisfies $\left(k+1\right)^{k+1}=4k^2$. For example, if $\kappa_{\max}=2$, then $\Delta t=\ln{\frac{32}{11}}$, while if $\kappa_{\max}>>1$, then $\Delta t\approx t_1\approx \frac{2\ln{2}}{\kappa_{\max}}$. The achievable minimal loss is illustrated in FIG. \ref{fig:noiseless_loss}.
\begin{equation}
b_{out}\left(0\leq t\leq t_1\right) =\frac{2\kappa_{1,\max}}{\kappa_{1,\max}-1}e^{-\kappa_{\max}t/2}-\frac{\kappa_{1,\max}+1}{\kappa_{1,\max}-1}e^{-t/2}
\end{equation}
\begin{equation}
\kappa_1\left(t>t_1\right)=\left(\frac{b_{in}\left(t\right)}{a_2\left(t\right) }\right)^2,\kappa_2\left(t\geq \Delta t\right)=\left(\frac{b'_{in}\left(t\right) }{a_2\left(t\right) }\right)^2
\end{equation}
\begin{equation}
a^2_2\left(t>\Delta t\right) =a^2\left(\Delta t\right)+\int_{\Delta t}^t\left(b_{in}^2\left(t'\right)+b'^2_{in}\left(t'\right)\right)dt'
\end{equation}
Thus, in the asymptotic limit, the two-pass method eliminates the loss of \(O(\gamma/\kappa_{\max})\) as long as \(\kappa_{\max}/\gamma>k \) in the noiseless regime, demonstrated by FIG. \ref{fig:noiseless_pulse}. 
\section{Exponential decay packet with intrinsic loss}
\begin{figure}[]
\begin{subfigure}[b]{0.35\textwidth}
\includegraphics[width=\textwidth]{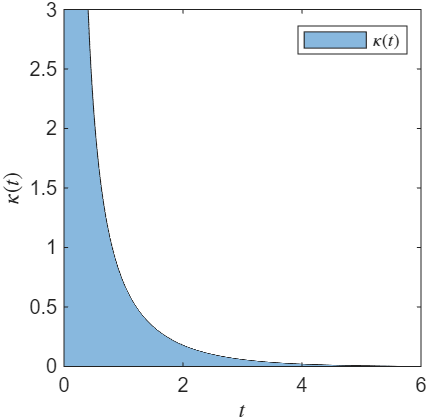}
\caption{Single-pass \(\kappa(t)\)}
\end{subfigure}
\begin{subfigure}[b]{0.35\textwidth}
\includegraphics[width=\textwidth]{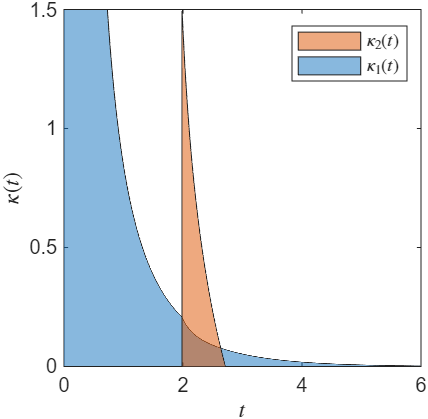}
\caption{Two-pass \(\kappa(t)\)}
\end{subfigure}
\begin{subfigure}[b]{0.35\textwidth}
\includegraphics[width=\textwidth]{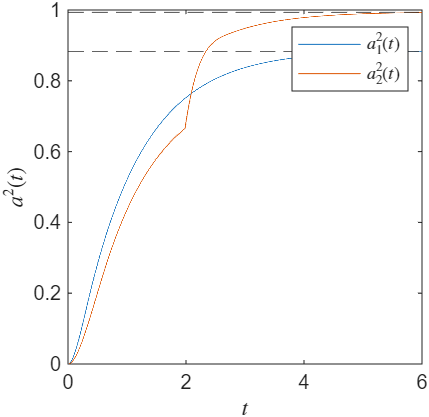}
\caption{\(a^2(t)\) for both schemes}
\end{subfigure}
\caption{Example pulse of single-pass with \(\kappa_{\max}=3\gamma\) and two-pass capture with \(\kappa_{1,\max}=\kappa_{2,\max}=1.5\gamma\), and the corresponding efficiency \(a^2(t)\), with intrinsic loss \(\kappa_i=0.001\gamma\). Dashed lines indicates the achievable maximum efficiencies of the two schemes. \(\kappa(t),\kappa_{1}(t)\) and \(\kappa_{2}(t)\) are in the units of \(\gamma\), while \(t\) is in the units of \(\gamma^{-1}\). }
\label{fig:noisy_pulse}
\end{figure}
Intrinsic loss \(\kappa_i\) models dissipation in the cavity and delay line, reducing capture efficiency. For simplicity, assume $\kappa_{i,cav}=\kappa_{i,delay}=\kappa_i\ll\kappa_{1,\max},\kappa_{2,\max},\gamma$. (If $\kappa_{i,cav}>\kappa_{i,delay}$, the solution would involve actively sending amplitude into the delay line; conversely, if $\kappa_{i,cav}<\kappa_{i,delay}$, the solution would involve premature second capture.) In such case, the zero reflection still provides the optimal solution, except both single and double capture will have an optimal stop time. Similar to the previous section, we choose time unit that satisfies \(\gamma=1\), i.e. \(t\) will be in the units of \(\gamma^{-1}\) and \(\kappa\) will be in the units of \(\gamma\), and let $\alpha=\frac{1-\kappa_i}{\kappa_{1,\max}}$. For the initial stage of the capture:
\begin{equation}
a(t\leq t_1)=\frac{2e^{-t/2}}{(1-\alpha)\sqrt{\kappa_{1,\max}}}\left(1-e^{-(1-\alpha)\kappa_{1,\max}t/2}\right)
\end{equation}
After \(t>t_1=\frac{2}{(1-\alpha)\kappa_{1,\max}}\ln{\frac{2}{1+\alpha}}\), the reflection can be eliminated, resulting in 
\begin{equation}
a_1(t\geq t_1)=\left(\alpha\kappa_{1,\max}\right)^{-1/2}e^{-t/2}\sqrt{\left(1+\alpha\right)e^{\alpha\kappa_{1,\max}(t-t_1)}-1}
\end{equation}We can now solve for maximal capture efficiency for the single-pass capture, which occurs at
\begin{align}
T_1&=t_1-\frac{\ln{(\kappa_i(1+\alpha))}}{\alpha\kappa_{1,\max}}\\
&=\frac{2\alpha\ln{2}-(1-\alpha)\ln\kappa_i-(1+\alpha)\ln{(1+\alpha)}}{\alpha\kappa_{1,\max}\left(1-\alpha\right)}
\end{align}
with the final efficiency of
\begin{equation}
a_1^2(T_1)=\kappa_i^{\frac{\kappa_i}{1-\kappa_i}}2^{-\frac{2}{\kappa_{1,\max}+\kappa_i-1}}(1+\alpha)^{\frac{1+\alpha}{\alpha\kappa_{1,\max}\left(1-\alpha\right)}}
\end{equation}
For the two-pass capture, once again assuming \(\Delta t\geq t_1\), we delay the recapture of reflection by
\begin{equation}
\Delta t=t_1+(\alpha\kappa_{1,\max})^{-1}\ln{\frac{1}{1+\alpha-\alpha\kappa_{1,\max}\kappa_{2,\max}^{-1}e^{\alpha\kappa_{1,\max}t_1}}}
\end{equation}
such that
\begin{equation}
\sqrt{\kappa_{2,\max}}a(\Delta t)=b'_{in}(\Delta t)=e^{-\kappa_i\Delta t/2}
\end{equation}
Similar to the previous section,
\begin{equation}
\frac{\kappa_{2,\max}}{\alpha\kappa_{1,\max}}\left(\frac{\left(1+\alpha\right)^{\alpha^{-1}+1}}{4}\right)^{\left(\alpha^{-1}-1\right)^{-1}}>1
\end{equation}
is required for 0 reflection at \(\Delta t\), which reduces to $\alpha^{-1}>k\approx 1.2834$ if \(\kappa_{1,\max}=\kappa_{2,\max}=\kappa_{\max}\). Applying this assumption, \(\Delta t\) reduces to
\begin{equation}
\Delta t=t_1+(\alpha\kappa_{1,\max})^{-1}\ln{\frac{1}{1+\alpha-\alpha e^{\alpha\kappa_{1,\max}t_1}}}
\end{equation}
and the capture efficiency is
\begin{equation}
\begin{split}
a_2^2(t)&=\frac{e^{-\kappa_it}-e^{-t}}{1-\kappa_i}\\&-\int_{t-\Delta t}^{t_1}\left(e^{-t'/2}-\sqrt{\kappa_{1,\max}}a(t')\right)^2e^{-\kappa_i(t-t')}dt'
\end{split}
\end{equation}
for \(\Delta t\leq t\leq \Delta t+t_1\), and
\begin{equation}
a_2^2(t)=\frac{e^{-\kappa_it}-e^{-t}}{1-\kappa_i}
\end{equation}
if $t\geq\Delta t+t_1$. 
Hence,
\begin{equation}
T_2=-\frac{\ln\kappa_i}{\alpha\kappa_{1,\max}},
\end{equation}
resulting with
\begin{equation}
a_2^2(T_2)=\kappa_i^{\frac{\kappa_i}{1-\kappa_i}}
\end{equation}
Note that we assumed $T_2>\Delta t+t_1$, which requires
\begin{equation}
\kappa_i\leq(1+\alpha)\left(\frac{1+\alpha}{2}\right)^{4\alpha/(1-\alpha)}-\frac{\alpha\kappa_{1,\max}}{\kappa_{2,\max}}\left(\frac{1+\alpha}{2}\right)^{2\alpha/(1-\alpha)}.
\end{equation}
For example, if $\kappa_{\max}=2$, $\kappa_i\lesssim0.225$. The two-capture process eliminated a factor of $a_1(T_1)/a_2(T_2)=\left(2^{-(1-\alpha)^{-1}}(1+\alpha)^{(1-\alpha)^{-1}+1/(2\alpha)}\right)^{\kappa_{1,\max}^{-1}}$, which is approximately \(1-\frac{\gamma}{\kappa_{1,\max}}\ln\frac{2}{\sqrt{e}}\) for small $\alpha$, and now the capture efficiency is now \(1-\widetilde{O}(\frac{\kappa_i}{\gamma})\), only limited by the ratio between intrinsic loss and decay rate of the input pulse, demonstrated in FIG. \ref{fig:noisy_pulse}. The overall capture time is also slightly shortened.
\begin{figure}[]
\begin{subfigure}[b]{0.45\textwidth}
\includegraphics[width=\textwidth]{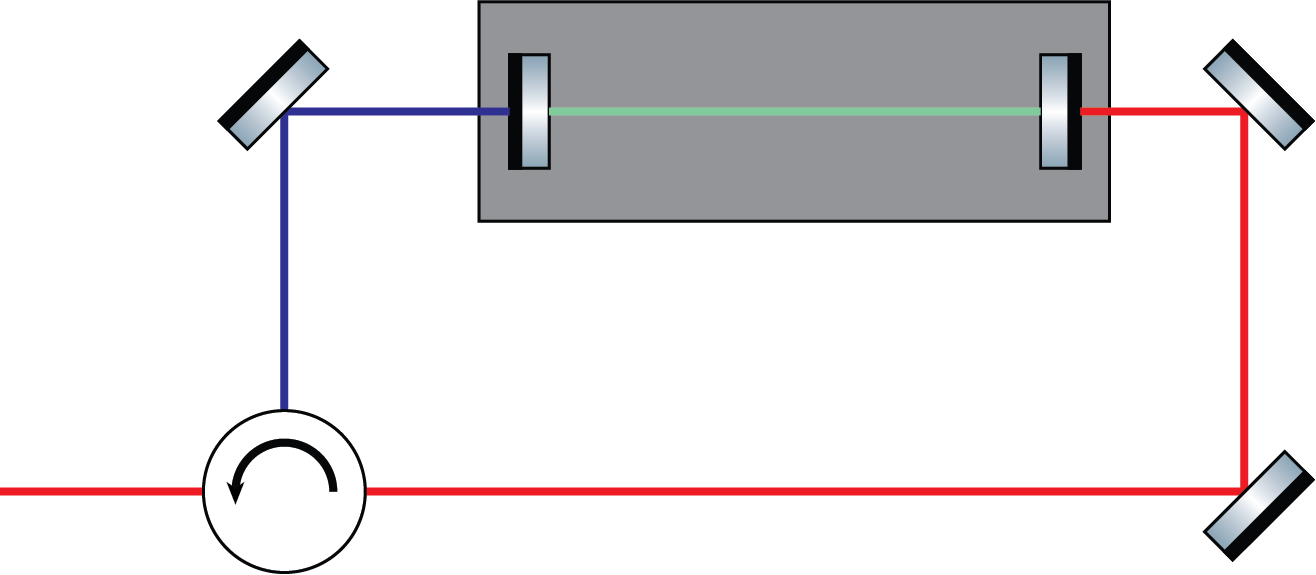}
\caption{Circulator-based setup}\label{fig:circ}
\end{subfigure}
\begin{subfigure}[b]{0.45\textwidth}
\includegraphics[width=\textwidth]{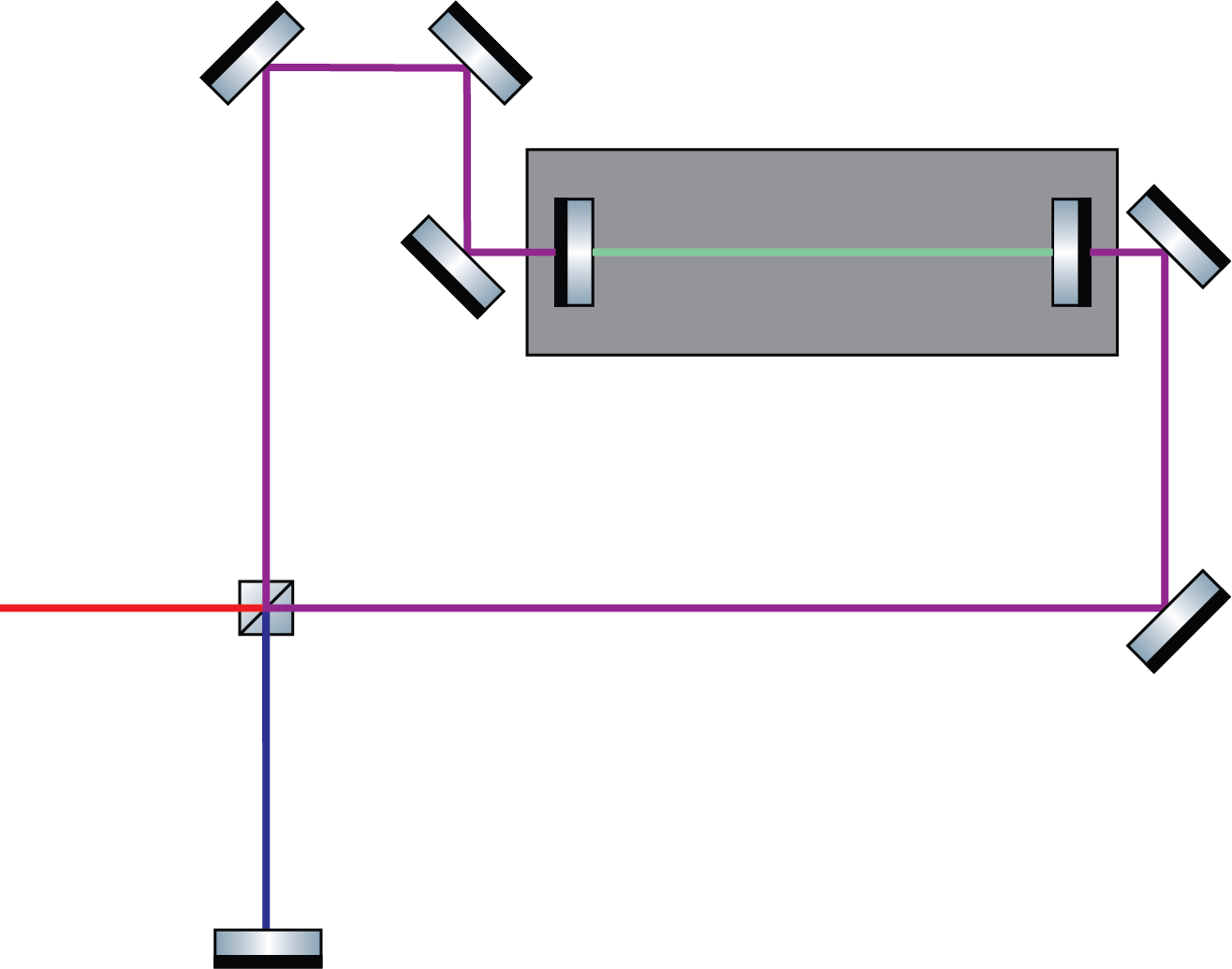}
\caption{Michelson-like setup}\label{fig:michelson}
\end{subfigure}
\begin{subfigure}[b]{0.45\textwidth}
\includegraphics[width=\textwidth]{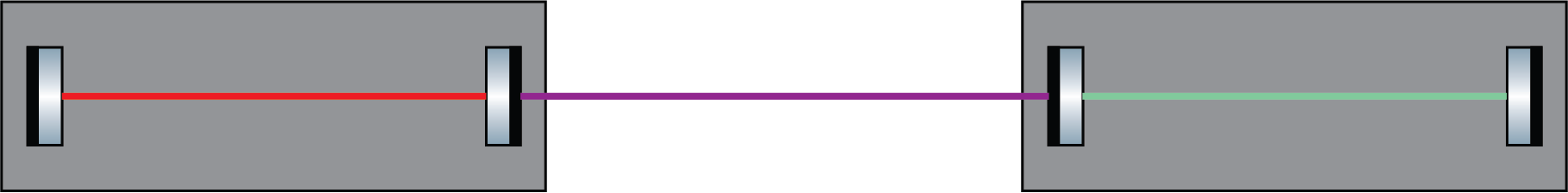}
\caption{Back-and-forth setup}\label{fig:twocav}
\end{subfigure}
\caption{Various setups for achieve two-pass capture. In \ref{fig:circ}, the reflection from the right port is redirected to the left port via a circulator, which can be implemented with non-reciprical optical components, such as a Faraday isolator. In \ref{fig:michelson}, two branches differs by \(\frac{\pi}{2}\), hence when the reflection reaches the beamsplitter, one branch accumulates an extra \mbox{\(\pi\)} phase, and the reflection will reach the mirror and be redirected to the cavity again instead of the source. In \ref{fig:twocav}, the reflection is utilized by the source cavity to extract the remaining excitations, then back to the destination cavity for the second-pass capture. Any reabsorption by the source cavity will be reemitted back to the destination cavity.}
\label{fig:setups}
\end{figure}
\section{Discussion}
The two-pass method significantly enhances photon capture over single-pass approaches by leveraging reflected pulses. Unlike single-pass methods, our two-pass method achieves perfect transfer for arbitrary pulses when \(\frac{\kappa_{\max}}{\max_t b_{in}(t)} \geq 2\ln 2 \) in the noiseless regime, and eliminates the \(e^{-O(\gamma/\kappa_{\max})}\) factor in capture efficiency in the noisy regime. Compared to prior single-pass techniques, our approach offers higher fidelity in realistic noisy conditions, providing potential improvements to quantum repeaters and memory nodes in quantum networks.

Current systems, both in optical and microwave regimes, are capable of shaping and capturing traveling photonic packets of timescales much shorter than the cavity intrinsic decays~\cite{Axline_2018,Yan_2022,Ritter_2012}. By applying our scheme, the reflection loss can be further reduced. There are many possible setups (refer to FIG. \ref{fig:setups}) to redirect the reflection back to the cavity in an orthogonal mode, including (a) optical circulator, (b) Michelson-like setup, and depending on the setup one can even just (c) utilize the reflection to extract the remaining excitation from the source. By reversing the process, our method can also be used to release arbitrarily shaped pulses, enhancing its utility for quantum pulse shaping. For quantum state transfer, our scheme can thus reduce both residual and reflection loss. Future work could explore thermal noise, non-linearity, alternative initial cavity state, designs for experimental implementations and multimode extensions.
\section{Conclusion}
Our two-pass method surpasses single-pass limitations, achieving perfect capture of arbitrary pulses in ideal conditions and superior fidelity with noise, advancing photonic manipulation for quantum technologies. These findings advance the toolkit for photonic state manipulation in quantum information systems.
\section{Acknowledgments}
We acknowledge support from the ARO(W911NF-23-1-0077), ARO MURI (W911NF-21-1-0325), AFOSR MURI (FA9550-21-1-0209, FA9550-23-1-0338), NSF (ERC-1941583, OMA-2137642, OSI-2326767, CCF-2312755, OSI-2426975), and the Packard Foundation (2020-71479).
\bibliography{savedrecs}
\end{document}